# Efficient single-cycle pulse compression of an ytterbium fiber laser at 10 MHz repetition rate

F. Köttig,[†,*] D. Schade,[†] J. R. Koehler, P. St.J. Russell, and F. Tani

*Max Planck Institute for the Science of Light, Staudtstr. 2, 91058 Erlangen, Germany*
[†]*These authors contributed equally to this work.*
[*]*Corresponding author:* felix.koettig@mpl.mpg.de

**Over the past years, ultrafast lasers with average powers in the 100 W range have become a mature technology, with a multitude of applications in science and technology. Nonlinear temporal compression of these lasers to few- or even single-cycle duration is often essential, yet still hard to achieve, in particular at high repetition rates. Here we report a two-stage system for compressing pulses from a 1030 nm ytterbium fiber laser to single-cycle durations with 5 μJ output pulse energy at 9.6 MHz repetition rate. In the first stage, the laser pulses are compressed from 340 to 25 fs by spectral broadening in a krypton-filled single-ring photonic crystal fiber (SR-PCF), subsequent phase compensation being achieved with chirped mirrors. In the second stage, the pulses are further compressed to single-cycle duration by soliton-effect self-compression in a neon-filled SR-PCF. We estimate a pulse duration of ~3.4 fs at the fiber output by numerically back-propagating the measured pulses. Finally, we directly measured a pulse duration of 3.8 fs (1.25 optical cycles) after compensating (using chirped mirrors) the dispersion introduced by the optical elements after the fiber, more than 50% of the total pulse energy being in the main peak. The system can produce compressed pulses with peak powers >0.6 GW and a total transmission exceeding 70%.**

## 1. INTRODUCTION

Extremely short low-noise laser pulse trains are widely used in science and technology, for example in ultrafast spectroscopy [1], attosecond science [2], light field-driven electron dynamics [3] and the generation of phase-stable mid-infrared pulses via intra-pulse difference frequency generation [4]. Despite advances in laser technology, generation of pulses with durations below a few cycles usually requires external nonlinear compression, which can be challenging, particularly at high repetition rates. While gas-filled hollow capillary fibers are routinely used for single-cycle pulse compression of high-energy (mJ) lasers at kHz repetition rates [5], scaling to lower pulse energies (μJ) and higher repetition rates (MHz) is difficult because the fiber loss scales with the inverse cube of the core size. Recently, an efficient compression scheme based on Heriott-type multipass cells was introduced [6], suitable for pulse compression down to ~20 fs at μJ energies and >10 MHz repetition rate [7]. However, compression to even shorter durations is currently limited by the residual dispersion of the intra-cavity multilayer coatings, which can reach significant values in multi-pass configurations. An alternative is single-pass spectral broadening in thin glass plates, followed by phase compensation [8]. Avoiding strong spatio-temporal distortions requires the compression ratio to be kept low, with the consequence that very short pump pulses are required if few- or single-cycle durations are to be reached [9,10].

Gas-filled hollow-core photonic crystal fiber (PCF) overcomes these limitations, offering an effective alternative solution for single-pass spectral broadening and pulse compression at μJ energies and MHz repetition rates. Among the different hollow-core PCF designs, anti-resonant-reflecting kagomé-type and single-ring PCFs (SR-PCFs) provide exceptionally broadband low-loss guidance (<10 dB km$^{-1}$ [11,12]) and a high damage threshold, making them ideal for high-power applications. These fibers offer meter-long interaction lengths at high intensities, which is ideal for exploiting nonlinear dynamics, while greatly relaxing the requirements on input pulse duration and energy. At the same time, the gas-filled PCF offers pressure-tunable anomalous dispersion, giving access to soliton dynamics which, in contrast to most other pulse compression schemes, enables direct temporal self-compression without need for additional dispersion compensation. In this way, compression to single-cycle pulses has been demonstrated at repetition rates up to 800 kHz, using optical parametric amplifiers as pump sources [13–15]. The great design flexibility of hollow-core PCF-based systems means that they can also be used with much less complex pump sources that provide longer pulses (≫100 fs), such as high-power high-repetition rate fiber and thin-disk lasers [16–18]. Few-cycle pulses have been generated using such systems, for example, centered at 1 μm wavelength at 38 MHz repetition rate [19], or at 2 μm at 1.25 MHz [20]. To the best of our knowledge, however, compression to single-cycle durations has not yet been demonstrated with these systems.

Here we report temporal compression to single-cycle durations of pulses from an ytterbium fiber laser. Starting with pulses >320 fs we compress to <3.8 fs at repetition rates up to 9.6 MHz with 5 μJ output pulse energy and >70% total transmission. The system is based on the two-stage design previously developed for generation of high-energy ultraviolet pulses [18], now adapted for high-repetition rate pulse compression (Fig. 1). The pump laser is a commercial 1030 nm ytterbium fiber laser (Active Fiber Systems GmbH) which delivers up to 100 W average power at repetition rates between 50 kHz and 19.2 MHz. The two compression stages are based on anti-resonant SR-PCFs, designed for normal-dispersion spectral broadening in the first stage and soliton-effect self-compression in the second stage. In anti-resonant PCFs, broadband low-loss guidance windows are interspersed with spectral anti-crossings between the fundamental core mode and core wall resonances, which strongly alter the dispersion and introduce high loss. This can impair the quality of ultrafast nonlinear dynamics, for example, pulse compression and ultraviolet light generation [21]. As a result, particular attention must be paid to the capillary wall thickness when designing the fiber structure, as described in sections 2 and 3 of this paper—it turns out that the requirements are well within reach of state-of-the-art fiber drawing techniques. Using commercially available pump lasers and specially designed hollow-core PCF, efficient pulse compression to single-cycle durations can be readily achieved and used in a plethora of different applications.

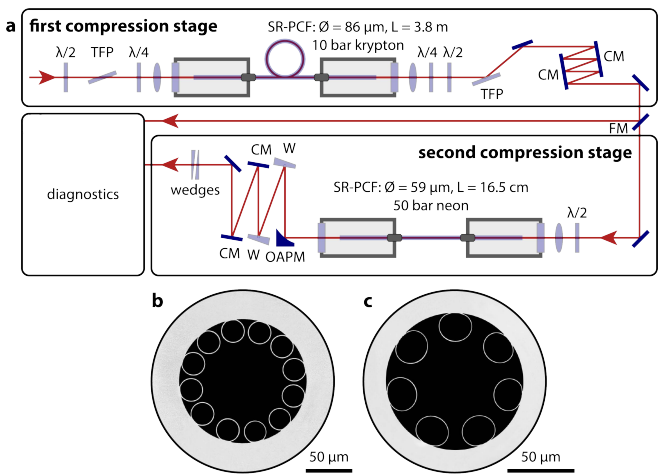

**Fig. 1.** Experimental setup. **(a)** Schematic of the experimental setup. λ/2: half-wave plate. λ/4: quarter-wave plate. TFP: thin-film polarizer. CM: negatively chirped mirror. FM: flip mirror. OAPM: off-axis parabolic mirror. W: glass wedge. **(b),(c)** Scanning electron micrographs of the SR-PCFs, core diameters 86 µm in the first compression stage and 59 µm in the second.

## 2. PULSE COMPRESSION FROM >320 fs to 25 fs

The pump laser produces pulses of duration >320 fs, which is too long to allow clean, high efficiency compression to a single-cycle in a single PCF stage. The pulses can, however, be compressed to 25 fs in a first PCF stage and then to <4 fs in a second stage. This two-stage design has the advantage that it provides 25 fs pulses useful in many other applications, for example generation of ultraviolet or mid-infrared pulses at high repetition rates [18,22,23]. In the experiment reported here, the pump laser was operated at repetition rates of 1, 4.8 and 9.6 MHz. Since the duration and shape of the laser pulses depends weakly on repetition rate, varying between 320 and 340 fs, energies between 7.4 and 7.55 µJ were used. The first compression stage is based on spectral broadening by self-phase modulation (SPM) in a krypton-filled SR-PCF, followed by phase compensation with negatively chirped mirrors. Since high intensities at high repetition rates can lead to instabilities at the fiber input, we used a long length (3.8 m) of 12-capillary SR-PCF with a large (86 µm) core diameter filled with 10 bar of krypton. These parameters reduced the required peak intensity to less than $10^{12}$ W cm$^{-2}$, allowing repetition rate scaling to at least 9.6 MHz. Additionally, the large core shifts the pump pulses into the normal dispersion region, preventing the onset of modulational instability, which would otherwise degrade the coherence of the compressed pulses. With a capillary wall thickness of ~730 nm, the fiber is pumped between the first- and second-order anti-crossings at ~1520 and ~770 nm, far enough away from 1030 nm to allow low-loss pulse propagation without any detrimental effects [21]. In the setup the fiber was coiled up in one turn with 0.8 m diameter, and a quarter-wave plate was placed before the fiber input to allow the polarization state to be adjusted to circular, so as to suppress nonlinear ellipse rotation along the fiber. After the fiber, a second quarter-wave plate changes the polarization back to linear, and a combination of broadband half-wave plate (Altechna) and thin-film polarizer (LAYERTEC GmbH, reflecting only vertical polarization) is used for power control. For temporal compression of the SPM-broadened pulses, negatively chirped mirrors (UltraFast Innovations GmbH) introduce a total group delay dispersion of $-3150$ fs$^2$ (21 bounces of $-150$ fs$^2$ each) to compensate for the positive chirp introduced by SPM in the fiber and the optics before the second compression stage. The optimal dispersion balance between SPM (positive chirp, decreasing with increasing spectral bandwidth) and the optics (net-negative chirp) is then fine-tuned by optimizing the input energy to the fiber while monitoring the intensity autocorrelation of the compressed pulses. The total transmission of the first compression stage is ~85%, with losses coming mainly from the optics, whereas the fiber transmission including incoupling is >96%.

For temporal characterization of the pulses, we used a home-built all-reflective dispersion-free second-harmonic generation frequency-resolved optical gating (SHG FROG) device in noncollinear geometry. The nonlinear medium was a 10-µm-thick beta barium borate crystal cut for type I phase matching at 800 nm (EKSMA Optics). Before retrieval, the measured FROG traces were adjusted to correct for the frequency-dependent SHG efficiency, including phase matching [24] as well as the frequency-dependent reflectivity of the optics after the crystal and the spectrometer calibration. For retrieval, we used the extended ptychographical iterative engine [25], the fundamental pulse spectrum retrieved from the measured FROG trace being used as an initial guess [26]. The retrieved pulse characteristics were then corrected by taking into account the frequency-dependent reflectivity of the optics before the crystal in the FROG device. With such rigorous measurement and retrieval, even sub-4 fs pulses could be reliably measured using SHG FROG, as reported in section 3.

Figure 2 shows FROG measurements of the pulses after the first compression stage. Since the pulses are negatively chirped, we numerically forward-propagated them to the fiber input to account for the dispersion of the optics before the input of the second fiber. At this location, the full-width at half-maximum (FWHM) pulse duration is 25 fs and the pulse energy is 6.3 µJ for all repetition rates, the fraction of energy in the main pulse lying between 63 and 67%. As can be seen from the FROG trace in Fig. 2, a long-duration low-intensity pedestal, already present in the pump laser pulses, is seen at delays greater than ±400 fs. We estimate that ~15% of the pump laser pulse energy is contained in this pedestal. Using amplitude and phase shaping techniques before amplification [27], we would be able to reduce this pedestal and increase the energy fraction in the main compressed pulse to >70%. Note that the first compression stage could be replaced by other schemes, for example a Heriott-type multipass cell, providing similar performance [28].

## 3. SINGLE-CYCLE PULSE COMPRESSION

Further pulse compression by SPM-based spectral broadening would require precise compensation of the nonlinear spectral phase, which becomes increasingly difficult as the pulse duration approaches the single-cycle regime. Soliton-effect self-compression, the dynamic interplay between spectral broadening and anomalous dispersion, provides an efficient alternative, inherently generating nearly chirp-free compressed pulses. Additionally, waveguide-based spectral broadening is largely free of spatio-temporal couplings, which have to be carefully managed in bulk solid-state systems [8–10]. For these reasons, we implemented a second compression stage based on soliton-effect self-compression in a 7-capillary SR-PCF with 59 µm core diameter.

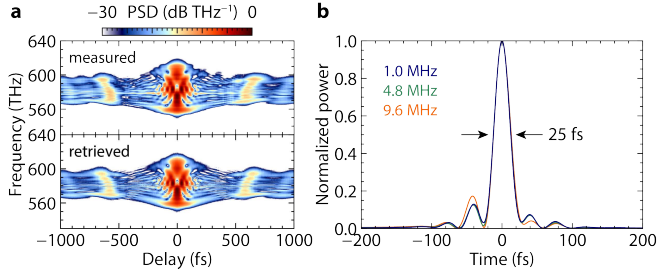

**Fig. 2.** First-stage pulse compression. **(a)** Measured and retrieved SHG FROG traces at 9.6 MHz repetition rate. **(b)** Retrieved pulses at repetition rates of 1, 4.8 and 9.6 MHz, numerically forward-propagated to the input of the second fiber. The input pulse energy was between 7.55 µJ (1 MHz) and 7.4 µJ (4.8 and 9.6 MHz) and the output energy was 6.3 µJ for all repetition rates, corresponding to a transmission of ~85%.

Because it is crucial at high repetition rates to reduce ionization as much as possible [18,29,30], we filled the fiber with 50 bar neon, avoiding using argon or krypton, which would provide similar dispersion and nonlinearity at much lower pressure, but have significantly lower ionization potentials. The capillary wall thickness of the fiber is 260 nm, placing the first-order anti-crossing at ~540 THz (555 nm), which is sufficiently far from the pump frequency so as not to impair pulse compression [21]. The fiber length (16.5 cm) was chosen so that pulse compression to <4 fs is reached directly at the fiber output for an input energy of ~6 µJ. A half-wave plate before the fiber was used to rotate the polarization to horizontal at the fiber output, as required for the subsequent optics and diagnostics. The output beam from the fiber was collimated with an aluminum-coated off-axis parabolic mirror (Newport Corporation), which initially had a rather low measured reflectivity of 92.6% for the compressed pulses and degraded further during the measurements, in particular at high average power. Using a parabolic mirror with higher reflectivity (for example silver-coated with typically >98% reflectivity) would mitigate this problem.

To understand the dynamics of pulse compression, the system was numerically modelled using a single-mode radially-resolved unidirectional full-field pulse propagation equation [31]. As input we used the pulse measured by FROG at 1 MHz repetition rate and numerically forward-propagated to the fiber input (Fig. 2), with an energy of 6 µJ. Because anti-crossings were far away from the pump frequency, the dispersion of the fiber could be approximated by a simple capillary model [32]. Literature values were taken for the dispersion [33] and the nonlinearity of the gas [34], and photoionization was included using the model from [35], with the Perelomov, Popov, Terent'ev ionization rates [36], modified with the Ammosov, Delone, Krainov coefficients [37].

Figure 3 shows the simulated pulse propagation in the fiber. For an input energy of 6 µJ, the soliton order is $N \sim 4.7$. Since the pump pulse lies in the anomalous dispersion region (the group velocity dispersion at the pump frequency is approximately $-270$ fs$^2$ m$^{-1}$ and the zero-dispersion frequency is ~430 THz), it undergoes soliton-effect self-compression to a FWHM duration as short as 2.5 fs at the fiber output. The choice of neon as filling gas meant that photoionization was negligible. The low peak plasma density (~3×10$^{12}$ cm$^{-3}$ on-axis) allowed repetition rate scaling over a wide range, as the energy lost due to ionization is only ~3 pJ per pulse, which minimizes gas heating that can cause a build-up of refractive index changes over time [30]. At the fiber output, the pulse has not yet reached its maximum temporal compression, so that compensation of the residual group delay dispersion (0.8 fs$^2$) would readily compress the pulse to 1.2 fs. Alternatively, the input energy or the fiber length could be increased to obtain even shorter pulses. Although compression to 1.2 fs has been conclusively predicted in hollow capillary fiber (which performs in a very similar manner to hollow-core PCF) by numerical back-propagation to the fiber output [38], the broader spectrum would exceed the bandwidth of our chirped mirrors and FROG device, preventing direct measurement of the compressed pulses. As can be seen from Figs. 4 and 5, the temporal pulse shape and spectrum at the fiber output are in good agreement with experimental measurements.

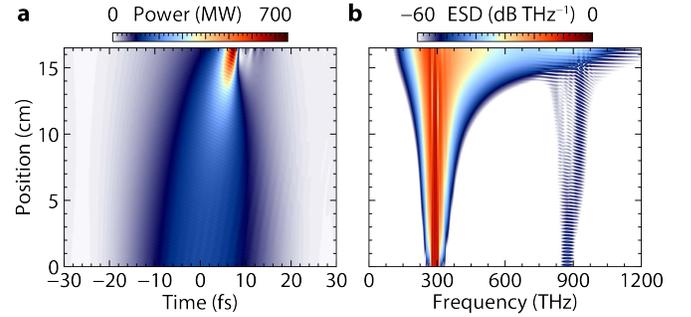

**Fig. 3.** Numerical simulation of the second-stage pulse compression. **(a)** Temporal and **(b)** spectral evolution of the pulse inside the fiber. At the fiber output, the pulse has compressed to a FWHM duration of 2.5 fs. The temporal pulse shape and spectrum are compared with measurements in Figs. 4 and 5.

In the experiments, pulse compression was initially optimized at 1 MHz repetition rate. For a launched energy of 5.9 µJ in the second fiber stage, the output spectrum closely matched the simulations (Fig. 5(d)), indicating pulse compression to <4 fs at the fiber output. Due to the slightly different pump pulses at other repetition rates, reaching the same pulse compression required pump energies of 5.5 µJ at 4.8 MHz and 6.3 µJ at 9.6 MHz. As a result, the output energy from the second stage (before the off-axis parabolic mirror) was 5.3 µJ at 1 MHz, 4.8 µJ at 4.8 MHz and 5.7 µJ at 9.6 MHz repetition rate, corresponding to a transmission of ~90%. Since Fresnel reflections at the uncoated 1.5-mm-thick magnesium fluoride output window of the gas cell (Korth Kristalle GmbH) introduced 5% loss, the fiber transmission including incoupling was ~95%. Figure 4 shows FROG measurements of the pulses after the second compression stage. Although the pulse is short directly at the fiber output, it is strongly dispersed by propagation through the gas cell, the output window and the air before reaching the FROG crystal. Numerical back-propagation of the measured pulses to the fiber output recovers well-compressed pulses with FWHM duration of 3.4 fs and 1.15 optical cycles. For such short pulses, rigorous estimates of the number of optical cycles within the FWHM of the pulse intensity envelope should be based on the phase evolution of the electric field, i.e., without assuming a carrier frequency (if the carrier is set to 1030 nm, the result is 1.0 optical cycles, an underestimate of the actual pulse width). Note that the uncertainty in the numerical back-propagation is of order ±1 fs and that the shortest pulse duration that can be obtained via numerical back-propagation, within the uncertainty of the optical elements, was 2.5 fs. Since our SHG FROG device cannot measure the entire spectral bandwidth of the pulses after the second fiber, the numerically back-propagated pulses are longer than in the simulations. Nevertheless, the measurements corroborate the conclusion that soliton-effect self-compression generates single-cycle pulses directly at the fiber output, without need for additional dispersion compensation.

To obtain compressed pulses at a target position for experiments requires compensation of the positive chirp introduced by all the optical elements after the fiber (even the beam path inside the gas cell and the beam path in air afterwards). To this end, we used a pair of double-angle negatively chirped mirrors (UltraFast Innovations GmbH, group delay dispersion of −80 fs$^2$ per pair) in combination with thin fused silica wedges (Altechna) for dispersion fine-tuning. The reflectivity of the chirped mirror pair, measured using the compressed pulses, is 93% (this is partially because spectral components at frequencies above ~530 THz (wavelengths below ~570 nm) are not reflected by the mirrors). The positive dispersion introduced by the fused silica wedges is relatively small (~1.5 mm material thickness) and could be easily eliminated by increasing the air path-length after the fiber (less than 2 m would be needed). In this way, additional loss could be avoided. Since only a small fraction of the 5 µJ can be used in the diagnostics, the signal was first attenuated before being sent to the chirped mirrors. As a result, only a small fraction of the second-stage output power was actually re-compressed for the FROG measurements. Since the chirped mirrors in the first compression stage are operated at full power, and such mirrors are routinely used at even higher pulse energy and average power (for example [39,40]), we are confident that re-compression of the full output power can easily be achieved if needed.

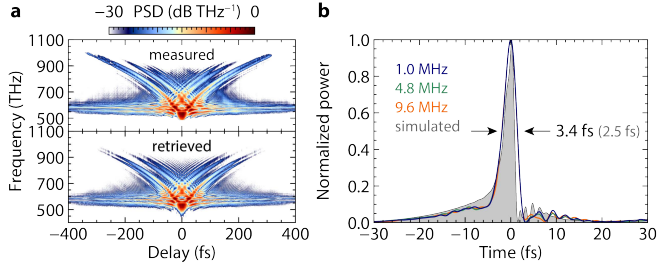

**Fig. 4.** Second-stage pulse compression, without dispersion compensation. **(a)** A zoom into the measured and retrieved SHG FROG traces at 9.6 MHz repetition rate over a ±400 fs time window. **(b)** Retrieved pulses at repetition rates of 1, 4.8 and 9.6 MHz, numerically backward-propagated to the fiber output. The simulated pulse (Fig. 3), with a duration of 2.5 fs, is shown under-shaded in gray. The output pulse energy before the off-axis parabolic mirror was 5.3 µJ (input energy 5.9 µJ) at 1 MHz repetition rate, 4.8 µJ (input 5.5 µJ) at 4.8 MHz and 5.7 µJ (input 6.3 µJ) at 9.6 MHz, corresponding to a transmission of ~90%.

Figure 5 shows FROG measurements of the pulses after the second compression stage, with dispersion compensation. As before, pulse compression was initially optimized at 1 MHz repetition rate by reproducing the reference output spectrum (Fig. 5(d)) and then adjusting wedges for the shortest pulse duration. The reproducibility of soliton self-compression in the second fiber meant that, if the reference spectrum was reproduced simply by tuning the energy launched into the second fiber, almost identical compressed pulses could be obtained at any other repetition rate, without need to re-optimize the dispersion compensation. With a FWHM pulse duration of 3.8 fs and 1.25 optical cycles, the pulses are close to their transform limit of ~3 fs. Despite the relatively long pulses from the pump laser (>320 fs), very high-quality compressed pulses were obtained, with more than 50% of the total pulse energy in the main peak and a pedestal less than 12% of the peak. We estimate that better pump laser pulses could further increase the energy in the main peak to >60%. As evident from the spectrogram in Fig. 5(c), the dispersion is not compensated at frequencies above ~470 THz (wavelengths below ~640 nm), which coincides with the specification of the chirped mirror design. Hence, dispersion compensation of the entire spectral bandwidth with optimized chirped mirrors could further compress the pulses. Finally, if a high-reflectivity off-axis parabolic mirror is used after the second stage, compressed pulses can be delivered with a total transmission >70% (including the chirped mirrors).

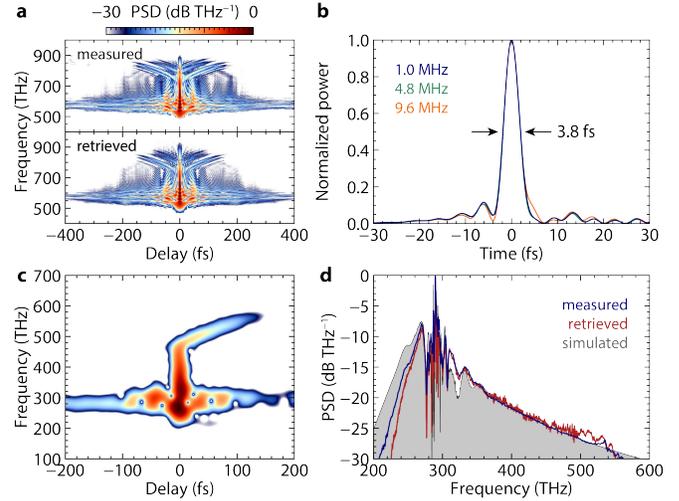

**Fig. 5.** Second-stage pulse compression, with dispersion compensation. **(a)** A zoom into the measured and retrieved SHG FROG traces at 9.6 MHz repetition rate over a ±400 fs time window. **(b)** Retrieved pulses at repetition rates of 1, 4.8 and 9.6 MHz. **(c)** Spectrogram of the pulse at 1 MHz repetition rate, calculated with a 15 fs Gaussian gate pulse. **(d)** Directly measured and FROG-retrieved spectrum of the pulse at 1 MHz repetition rate. The simulated spectrum (Fig. 3) is shown under-shaded in gray.

## 4. STABILITY MEASUREMENTS

Stable operation of high-power high-repetition rate systems can be prevented by various effects, in particular when precise optical alignment is necessary, for example, for coupling into fibers. Slow thermal drift and beam pointing fluctuations up to kHz-level frequencies can be compensated using beam stabilization systems. Additional noise sources at higher frequencies, introduced for example by nonlinear processes, are more difficult to suppress. In the first compression stage, coherent spectral broadening is achieved by SPM in the normal dispersion region and in the second stage, low-order ($N<5$) soliton-effect self-compression takes place. Both processes preserve coherence [41], as experimentally confirmed by the high-fidelity scanning-delay FROG measurements, each of which took ~10 min.

To further characterize the long-term stability of the system, we measured the output spectrum and power of the system at intervals of a second over 30 min (Fig. 6). At 1 MHz repetition rate, the spectrum changed only marginally and the power remained constant. At 9.6 MHz, on the other hand, we had to realign the incoupling to the second fiber stage after 10 and 20 min due to slow thermal drift of the output from the first fiber stage (realignment was done using two irises before the second fiber stage and took only a few minutes). After realignment, the output spectrum remained constant, whereas the power dropped by 2.6% after 30 min, from 46.6 W to 45.4 W. The major part of this drop comes from degradation of the aluminum-coated off-axis parabolic mirror used to collimate the output from the second fiber. Its reflectivity was measured to 92.6% before the experiment, compared to 91.5% afterwards, corresponding to a 1.2% drop in reflectivity, which we expect was even higher during the experiment when the mirror heated up strongly. This is supported by the stability of the output spectrum after 10 min, which was sensitive to energy fluctuations, but remained constant after realignment. If a high-reflectivity mirror is used after the second fiber, together with a beam stabilization system after the first fiber stage to correct slow beam pointing drifts (or a sufficiently long warm-up period), stable operation can be achieved over long periods without realignment, even at high repetition rates.

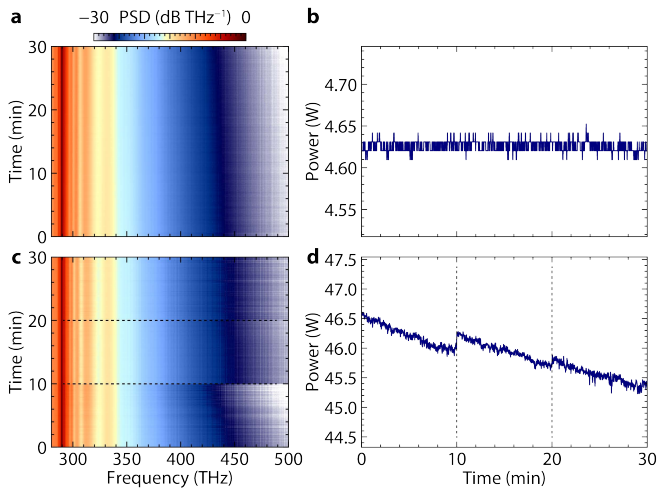

**Fig. 6.** Long-term stability of the system. **(a)** Output spectrum and **(b)** average power after the off-axis parabolic mirror of the second fiber stage at 1 MHz repetition rate, measured once a second over 30 min. **(c)** Output spectrum and **(d)** average power at 9.6 MHz repetition rate. Slow thermal drift meant that the second fiber stage had to be re-aligned after 10 and 20 min (indicated by the dashed lines). The decrease in output power is mainly due to damage to the off-axis parabolic mirror after the fiber (the reflectivity was 92.6% before the experiment and 91.5% afterwards).

In addition to the long-term stability, we measured the pulse energy fluctuations of the system on short timescales using a 150 MHz silicon photodiode and a radio-frequency spectrum analyzer (Agilent E4440A). Figure 7 shows the relative intensity noise (RIN) measured after the second fiber stage at 9.6 MHz and compared to the pump laser noise (similar results were obtained at the lower repetition rates). Most of the intensity noise is at frequencies >100 kHz, where the two-stage compression system introduces an additional ~0.1% integrated RIN. The largest contribution to the high-frequency intensity noise is from broad peaks that are already present in the RIN spectrum of the laser, which are amplified by the highly nonlinear compression in the two fiber stages. In the low-frequency range (<100 kHz), on the other hand, there is no significant additional intensity noise introduced by the compression system, which means that acoustic and mechanical noise sources do not strongly couple to the compression process and that stabilization of the slow beam-pointing drift at high average power (Fig. 6) is sufficient. In total, the integrated RIN values from 2 Hz to 4.8 MHz (half the repetition rate) are 0.3% for the laser and 0.4% for the compressed pulses, which compares well with low-noise thin-disk oscillators [16].

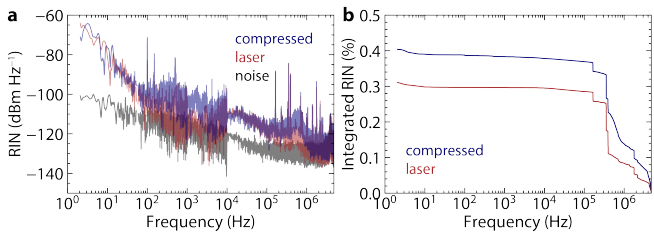

**Fig. 7.** Relative intensity noise (RIN) measured at 9.6 MHz repetition rate. **(a)** RIN of the laser and the output from the second fiber stage from 2 Hz to 4.8 MHz (half the repetition rate). The change in signal fluctuations at 1 and 10 kHz is caused by a change in the radio-frequency bandwidth of the spectrum analyzer. **(b)** Integrated RIN (normalized to the DC component before integration), starting from 4.8 MHz, plotted as function of the lower frequency limit.

## 5. CONCLUSIONS AND OUTLOOK

In conclusion, single-cycle pulses can be generated by two-stage nonlinear temporal compression of pulses from a 1030 nm ytterbium fiber laser at repetition rates up to 9.6 MHz. In the first stage, the pump laser pulses are compressed from >320 to 25 fs and in the second stage, from 25 to <3.8 fs (1.25 optical cycles) at 5 µJ output energy. The overall compression factor is 100, with a total transmission as high as >70%. After a 20 min warm-up time, the system provides stable low-noise operation with 0.4% integrated RIN (from 2 Hz to 4.8 MHz).

Since the second compression stage is based on soliton-effect self-compression, only the dispersion of the optics after the fiber must be compensated to deliver compressed pulses to a target location. This does not require complex adaptive phase shaping but only chirped mirrors and fused silica wedges. At the same time, the compression is highly reproducible simply by recovering the output spectrum at which the dispersion compensation was optimized. Once this spectrum is recovered, almost identical pulses are generated, without need to re-optimize the dispersion compensation.

Obtaining a precisely-defined single-cycle optical waveform requires fine control of the pulse carrier-envelope phase (CEP). Recently, it was shown that CEP-stable few-cycle pulses can be generated at 100 kHz repetition rate from a 300 µJ ytterbium fiber laser by two-stage nonlinear compression in gas-filled hollow capillary fibers [42]. It was also shown that soliton-effect self-compression in gas-filled PCF can be used to generate CEP-stable pulses in the single-cycle regime at 800 kHz repetition rate with only 1 µJ pump pulses from an optical parametric amplifier [15]. Implementing CEP-stabilization schemes as described in [42], we expect that phase-stable pulses can be generated also with our system. This would enable many exciting experiments, for example in phase-sensitive ultrafast spectroscopy and strong-field physics, with a simple setup at unprecedented repetition rate and power.

## REFERENCES

1. R. Berera, R. van Grondelle, and J. T. M. Kennis, "Ultrafast transient absorption spectroscopy: principles and application to photosynthetic systems," Photosynth. Res. **101,** 105 (2009).
2. F. Krausz, and M. Ivanov, "Attosecond physics," Rev. Mod. Phys. **81,** 163 (2009).
3. T. Higuchi, C. Heide, K. Ullmann, H. B. Weber, and P. Hommelhoff, "Light-field-driven currents in graphene," Nature **550,** 224 (2017).
4. I. Pupeza, D. Sánchez, J. Zhang, N. Lilienfein, M. Seidel, N. Karpowicz, T. Paasch-Colberg, I. Znakovskaya, M. Pescher, W. Schweinberger, V. Pervak, E. Fill, O. Pronin, Z. Wei, F. Krausz, A. Apolonski, and J. Biegert, "High-power sub-two-cycle mid-infrared pulses at 100 MHz repetition rate," Nat. Photonics **9,** 721 (2015).
5. M. Nisoli, S. De Silvestri, O. Svelto, R. Szipöcs, K. Ferencz, C. Spielmann, S. Sartania, and F. Krausz, "Compression of high-energy laser pulses below 5 fs," Opt. Lett. **22,** 522 (1997).
6. J. Schulte, T. Sartorius, J. Weitenberg, A. Vernaleken, and P. Russbueldt, "Nonlinear pulse compression in a multi-pass cell," Opt. Lett. **41,** 4511 (2016).
7. K. Fritsch, M. Poetzlberger, V. Pervak, J. Brons, and O. Pronin, "All-solid-state multipass spectral broadening to sub-20 fs," Opt. Lett. **43,** 4643 (2018).
8. C.-H. Lu, Y.-J. Tsou, H.-Y. Chen, B.-H. Chen, Y.-C. Cheng, S.-D. Yang, M.-C. Chen, C.-C. Hsu, and A. H. Kung, "Generation of intense supercontinuum in condensed media," Optica **1,** 400 (2014).
9. O. Pronin, M. Seidel, F. Lücking, J. Brons, E. Fedulova, M. Trubetskov, V. Pervak, A. Apolonski, T. Udem, and F. Krausz, "High-power multi-megahertz source of waveform-stabilized few-cycle light," Nat. Commun. **6,** 6988 (2015).
10. C.-H. Lu, T. Witting, A. Husakou, M. J.J. Vrakking, A. H. Kung, and F. J. Furch, "Sub-4 fs laser pulses at high average power and high repetition rate from an all-solid-state setup," Opt. Express **26,** 8941 (2018).
11. M. Maurel, M. Chafer, A. Amsanpally, M. Adnan, F. Amrani, B. Debord, L. Vincetti, F. Gérôme, and F. Benabid, "Optimized inhibited-coupling

11. Kagome fibers at Yb-Nd:Yag (8.5 dB/km) and Ti:Sa (30 dB/km) ranges," Opt. Lett. **43,** 1598 (2018).
12. B. Debord, A. Amsanpally, M. Chafer, A. Baz, M. Maurel, J. M. Blondy, E. Hugonnot, F. Scol, L. Vincetti, F. Gérôme, and F. Benabid, "Ultralow transmission loss in inhibited-coupling guiding hollow fibers," Optica **4,** 209 (2017).
13. T. Balciunas, C. Fourcade-Dutin, G. Fan, T. Witting, A. A. Voronin, A. M. Zheltikov, F. Gerome, G. G. Paulus, A. Baltuska, and F. Benabid, "A strong-field driver in the singlecycle regime based on self-compression in a kagome fibre," Nat. Commun. **6,** 6117 (2015).
14. U. Elu, M. Baudisch, H. Pires, F. Tani, M. H. Frosz, F. Köttig, A. Ermolov, P. St.J. Russell, and J. Biegert, "High average power and single-cycle pulses from a mid-IR optical parametric chirped pulse amplifier," Optica **4,** 1024 (2017).
15. A. Ermolov, C. Heide, P. Dienstbier, F. Köttig, F. Tani, P. Hommelhoff, and P. St.J. Russell, "Carrier-envelope-phase-stable soliton-based pulse compression to 4.4 fs and ultraviolet generation at the 800 kHz repetition rate," Opt. Lett. **44,** 5005 (2019).
16. F. Emaury, A. Diebold, C. J. Saraceno, and U. Keller, "Compact extreme ultraviolet source at megahertz pulse repetition rate with a low-noise ultrafast thin-disk laser oscillator," Optica **2,** 980 (2015).
17. S. Hädrich, M. Krebs, A. Hoffmann, A. Klenke, J. Rothhardt, J. Limpert, and A. Tünnermann, "Exploring new avenues in high repetition rate table-top coherent extreme ultraviolet sources," Light Sci. Appl. **4,** e320 (2015).
18. F. Köttig, F. Tani, C. Martens Biersach, J. C. Travers, and P. St.J. Russell, "Generation of microjoule pulses in the deep ultraviolet at megahertz repetition rates," Optica **4,** 1272 (2017).
19. K. F. Mak, M. Seidel, O. Pronin, M. H. Frosz, A. Abdolvand, V. Pervak, A. Apolonski, F. Krausz, J. C. Travers, and P. St. J. Russell, "Compressing μJ-level pulses from 250 fs to sub-10 fs at 38-MHz repetition rate using two gas-filled hollow-core photonic crystal fiber stages," Opt. Lett. **40,** 1238 (2015).
20. M. Gebhardt, C. Gaida, T. Heuermann, F. Stutzki, C. Jauregui, J. Antonio-Lopez, A. Schulzgen, R. Amezcua-Correa, J. Limpert, and A. Tünnermann, "Nonlinear pulse compression to 43 W GW-class few-cycle pulses at 2 μm wavelength," Opt. Lett. **42,** 4179 (2017).
21. F. Tani, F. Köttig, D. Novoa, R. Keding, and P. St.J. Russell, "Effect of anti-crossings with cladding resonances on ultrafast nonlinear dynamics in gas-filled photonic crystal fibers," Photon. Res. **6,** 84 (2018).
22. F. Köttig, F. Tani, P. Uebel, P. St.J. Russell, and J. C. Travers, "High Average-Power and Energy Deep-Ultraviolet Femtosecond Pulse Source Driven by 10 MHz Fibre-Laser," CLEO/Europe 2015, PD_A_7.
23. F. Köttig, D. Novoa, F. Tani, M. C. Günendi, M. Cassataro, J. C. Travers, and P. St.J. Russell, "Mid-infrared dispersive wave generation in gas-filled photonic crystal fibre by transient ionization-driven changes in dispersion," Nat. Commun. **8,** 813 (2017).
24. A. Baltuska, M. S. Pshenichnikov, and D. A. Wiersma, "Second-Harmonic Generation Frequency-Resolved Optical Gating in the Single-Cycle Regime," IEEE J. Quantum Electron. **35,** 459 (1999).
25. P. Sidorenko, O. Lahav, Z. Avnat, and O. Cohen, "Ptychographic reconstruction algorithm for frequency-resolved optical gating: super-resolution and supreme robustness," Optica **3,** 1320 (2016).
26. R. Jafari, T. Jones, and R. Trebino, "100% reliable algorithm for second-harmonic-generation frequency-resolved optical gating," Opt. Express **27,** 2112 (2019).
27. N. C. Becker, S. Hädrich, T. Eidam, F. Just, K. Osvay, Z. Várallyay, J. Limpert, A. Tünnermann, T. Pertsch, and F. Eilenberger, "Adaptive pre-amplification pulse shaping in a high-power, coherently combined fiber laser system," Opt. Lett. **42,** 3916 (2017).
28. G. Barbiero, R. N. Ahmad, H. Wang, F. Köttig, D. Novoa, F. Tani, J. Brons, P. St.J. Russell, F. Krausz, and H. Fattahi, "Towards 45 watt single-cycle pulses from Yb:YAG thin-disk oscillators," CLEO/Europe 2019, cf_1_2.
29. J. R. Koehler, F. Köttig, B. M. Trabold, F. Tani, and P. St.J. Russell, "Long-Lived Refractive-Index Changes Induced by Femtosecond Ionization in Gas-Filled Single-Ring Photonic-Crystal Fibers," Phys. Rev. Applied **10,** 064020 (2018).
30. R. Koehler, F. Köttig, D. Schade, F. Tani, and P. St.J. Russell, "Buildup of Post-Recombination Refractive Index Changes in Krypton Photoionized at High Repetition Rates," accepted HILAS 2020.
31. A. Couairon, E. Brambilla, T. Corti, D. Majus, O. de J. Ramírez-Góngora, and M. Kolesik, "Practitioner's guide to laser pulse propagation models and simulation," Eur. Phys. J. Spec. Top. **199,** 5 (2011).
32. J. Nold, P. Hölzer, N. Y. Joly, G. K. L. Wong, A. Nazarkin, A. Podlipensky, M. Scharrer, and P. St.J. Russell, "Pressure-controlled phase matching to third harmonic in Ar-filled hollow-core photonic crystal fiber," Opt. Lett. **35,** 2922 (2010).
33. A. Börzsönyi, Z. Heiner, M. P. Kalashnikov, A. P. Kovács, and K. Osvay, "Dispersion measurement of inert gases and gas mixtures at 800 nm," Appl. Opt. **47,** 4856 (2008).
34. H. J. Lehmeier, W. Leupacher, and A. Penzkofer, "Nonresonant Third Order Hyperpolarizability of Rare Gases and $N_2$ Determined by Third Harmonic Generation," Opt. Commun. **56,** 67 (1985).
35. M. Geissler, G. Tempea, A. Scrinzi, M. Schnürer, F. Krausz, and T. Brabec, "Light Propagation in Field-Ionizing Media: Extreme Nonlinear Optics," Phys. Rev. Lett. **83,** 2930 (1999).
36. A. M. Perelomov, V. S. Popov, and M. V. Terent'ev, "Ionization of Atoms in an Alternating Electric Field," Sov. Phys. JETP **23,** 924 (1966).
37. A. Couairon and A. Mysyrowicz, "Femtosecond filamentation in transparent media," Phys. Rep. **441,** 47 (2007).
38. J. C. Travers, T. F. Grigorova, C. Brahms, and F. Belli, "High-energy pulse self-compression and ultraviolet generation through soliton dynamics in hollow capillary fibres," Nat. Photonics **13,** 547 (2019).
39. S. Hädrich, M. Kienel, M. Müller, A. Klenke, J. Rothhardt, R. Klas, T. Gottschall, T. Eidam, A. Drozdy, P. Jójárt, Z. Várallyay, E. Cormier, K. Osvay, A. Tünnermann, and J. Limpert, "Energetic sub-2-cycle laser with 216 W average power," Opt. Lett. **41,** 4332 (2016).
40. T. Nagy, S. Hädrich, P. Simon, A. Blumenstein, N. Walther, R. Klas, J. Buldt, H. Stark, S. Breitkopf, P. Jójárt, I. Seres, Z. Várallyay, T. Eidam, and J. Limpert, "Generation of three-cycle multi-millijoule laser pulses at 318 W average power," Optica **6,** 1423 (2019).
41. J. M. Dudley, G. Genty, and S. Coen, "Supercontinuum generation in photonic crystal fiber," Rev. Mod. Phys. **78,** 1135 (2006).
42. E. Shestaev, D. Hoff, A. M. Sayler, A. Klenke, S. Hädrich, F. Just, T. Eidam, P. Jójárt, Z. Várallyay, K. Osvay, G. G. Paulus, A. Tünnermann, and J. Limpert, "High-power ytterbium-doped fiber laser delivering few-cycle, carrier-envelope phase-stable 100 μJ pulses at 100 kHz," Opt. Lett. **45,** 97 (2020).